\documentclass[%
reprint,
superscriptaddress,
amsmath,
amssymb,
aps,
pra,
longbibliography,
floatfix,
]{revtex4-2}

\usepackage{times}
\usepackage[colorlinks=true,linkcolor=blue,urlcolor=blue,citecolor=blue]{hyperref}
\usepackage{graphicx}
\usepackage{bm}
\usepackage{bbold}
\usepackage{dsfont}
\usepackage{xcolor}

\newcommand{\ti}{t_\mathrm{i}}
\newcommand{\tf}{t_\mathrm{f}}
\newcommand{\ket}[1]{\vert #1 \rangle}

\begin{document}
\title{Two-photon-transition superadiabatic passage in an nitrogen-vacancy center in diamond}

\author{Musang Gong}
\affiliation{School of Physics, Hubei Key Laboratory of Gravitation and Quantum Physics, Institute for Quantum Science and Engineering, Huazhong University of Science and Technology, Wuhan 430074, China}
\affiliation{International Joint Laboratory on Quantum Sensing and Quantum Metrology, Huazhong University of Science and Technology, Wuhan 430074, China}

\author{Min Yu}
\email{min\_yu@hust.edu.cn}
\affiliation{School of Physics, Hubei Key Laboratory of Gravitation and Quantum Physics, Institute for Quantum Science and Engineering, Huazhong University of Science and Technology, Wuhan 430074, China}
\affiliation{International Joint Laboratory on Quantum Sensing and Quantum Metrology, Huazhong University of Science and Technology, Wuhan 430074, China}

\author{Yaoming Chu}
\affiliation{School of Physics, Hubei Key Laboratory of Gravitation and Quantum Physics, Institute for Quantum Science and Engineering, Huazhong University of Science and Technology, Wuhan 430074, China}
\affiliation{International Joint Laboratory on Quantum Sensing and Quantum Metrology, Huazhong University of Science and Technology, Wuhan 430074, China}

\author{Wei Chen}
\affiliation{School of Physics, Hubei Key Laboratory of Gravitation and Quantum Physics, Institute for Quantum Science and Engineering, Huazhong University of Science and Technology, Wuhan 430074, China}
\affiliation{International Joint Laboratory on Quantum Sensing and Quantum Metrology, Huazhong University of Science and Technology, Wuhan 430074, China}

\author{Qingyun Cao}
\affiliation{Institut für Quantenoptik and IQST, Universität Ulm, Albert-Einstein-Allee 11, 89081 Ulm, Germany}
\affiliation{International Joint Laboratory on Quantum Sensing and Quantum Metrology, Huazhong University of Science and Technology, Wuhan 430074, China}

\author{Ning Wang}
\affiliation{School of Physics, Hubei Key Laboratory of Gravitation and Quantum Physics, Institute for Quantum Science and Engineering, Huazhong University of Science and Technology, Wuhan 430074, China}
\affiliation{International Joint Laboratory on Quantum Sensing and Quantum Metrology, Huazhong University of Science and Technology, Wuhan 430074, China}

\author{Jianming Cai}
\affiliation{School of Physics, Hubei Key Laboratory of Gravitation and Quantum Physics, Institute for Quantum Science and Engineering, Huazhong University of Science and Technology, Wuhan 430074, China}
\affiliation{International Joint Laboratory on Quantum Sensing and Quantum Metrology, Huazhong University of Science and Technology, Wuhan 430074, China}
\affiliation{Shanghai Key Laboratory of Magnetic Resonance, East China Normal University, Shanghai 200062, China}

\author{Ralf Betzholz}
\affiliation{School of Physics, Hubei Key Laboratory of Gravitation and Quantum Physics, Institute for Quantum Science and Engineering, Huazhong University of Science and Technology, Wuhan 430074, China}
\affiliation{International Joint Laboratory on Quantum Sensing and Quantum Metrology, Huazhong University of Science and Technology, Wuhan 430074, China}

\author{Luigi Giannelli}
\affiliation{Dipartimento di Fisica e Astronomia “Ettore Majorana”, Università di Catania, Via S. Sofia 64, 95123, Catania, Italy}
\affiliation{CNR-IMM, UoS Universit\`a, 95123, Catania, Italy}

\begin{abstract}
Reaching a given target quantum state with high fidelity and fast operation speed close to the quantum limit represents an important goal in quantum information science. Here, we experimentally demonstrate superadiabatic quantum driving to achieve population transfer in a three-level solid-state spin system. Starting from traditional stimulated Raman adiabatic passage (STIRAP), our approach implements superadiabatic corrections to the STIRAP Hamiltonians with several paradigmatic pulse shapes. It requires no need of intense microwave pulses or long transfer times and shows enhanced robustness over pulse imperfections. These results might provide a useful tool for quantum information processing and coherent manipulations of quantum systems.
\end{abstract}

\maketitle
\date{\today}

\section{Introduction}
The demand for accurate control of quantum systems is steadily increasing with the continuing development of modern quantum technologies, which include, among others, quantum information processing~\cite{farhi2001quantum,monroe2013scaling,chen2020parallel} and quantum sensing~\cite{kasevich2002coherence,kotru2015large,buckley2010spin}. The goal of quantum control is to reach a desired target state efficiently while being robust against experimental errors such as pulse imperfections and frequency detunings. A host of methods have been developed for this purpose, such as adiabatic-passages~\cite{shapiro2000coherent,rangelov2005stark,torosov2011high,kovachy2012adiabatic,di2015population,di2016coherent,vepsalainen2016quantum,siewert2009advanced,brown2021reinforcement}, superadiabatic techniques~\cite{lim1991superadiabatic,berry2009transitionless,demirplak2003adiabatic,demirplak2008consistency,demirplak2005assisted}, and jump protocols~\cite{gong2023accelerated,wang2016necessary,liu2022arxiv}. 

Among these methods, the so-called stimulated Raman adiabatic passage (STIRAP)~\cite{gaubatz1990population,vitanov2017stimulated} is an absolute gold standard. This paradigm protocol achieves high-fidelity transfer of population between two state, which are not directly coupled, using an intermediate state. It utilizes two coherent optical or microwave fields to transfer the population. First, the so-called Stokes field connecting the target state and the intermediate state is applied and only then the pump field connecting the initial state and the intermediate states is turned on. Although this sequence might seem counter-intuitive, when these pulses are applied sufficiently slowly, the system adiabatically follows an eigenstate of the Hamiltonian, reaching nearly perfect transfer of population from the initial state to the final state. Unfortunately, this method has some profound drawbacks. Since it is contingent on the fulfillment of the adiabatic condition, the resulting protocol duration may be rather long. Furthermore, it is not necessarily applicable in all quantum systems because relatively high pulse intensities might be required. 

One way to circumvent these drawbacks and nevertheless achieve high-efficiency population transfer are superadiabatic protocols, which are also referred to as transitionless-driving protocols. Such superadiabatic protocols, as a type of shortcut to adiabaticity, are able to suppress nondiabatic excitations while the system follows the instantaneous eigenstate of the Hamiltonian during the evolution. The main obstacle that has to be overcome in order to implement superadiabatic protocols is the implementation of complex couplings with externally controlled and stable phases~\cite{peierls1933theorie}, which would, for example, require lasers with low phase noise in optical setups.

In this work, we implement a two-photon-transition superadiabatic-passage protocol~\cite{giannelli2014three,kumar2016stimulated} in the three-level system formed by the electronic ground-state triplet of a single nitrogen-vacancy (NV) center in diamond~\cite{jelezko2006single,doherty2013nitrogen,gruber1997scanning,rondin2014magnetometry}. The NV center has been widely investigated during the last two decades due to its long coherence time at room temperature and well-developed coherent control techniques~\cite{jelezko2004observation,bar2013solid,balasubramanian2009ultralong,chu2015all,shu2018observation,tian2020quantum,cao2020protecting}, making it an ideal test bed for the implementation of new quantum technologies, such as quantum information processing~\cite{wrachtrup2006processing,dolde2013room,cai2013large,xiang2013hybrid,kurizki2015quantum,yu2020experimental,yu2021experimental,yu2022quantum}, quantum computing~\cite{neumann2008multipartite,zu2014experimental}, as well as quantum-sensing~\cite{kucsko2013nanometre,Acosta2010prl,neumann2013nl,plakhotnik2014nl,knauer2020situ,trusheim2016njp,giannelli2016laser,dolde2011electric,maze2008nanoscale,balasubramanian2008nanoscale,welter2022magnetometry}. The energy splittings between the spin states of the ground-state triplet are in the microwave regime, where we can overcome the above obstacle. In detail, we employ four different microwave pulses to demonstrate a transfer efficiency which is superior compared with the traditional STIRAP protocol, not only in terms of time constraints but also in terms of the robustness against pulse imperfections.

The paper is structured as follows. In Sec.~\ref{sec:protocols}, we give a basic summary of the theoretical description of both the traditional STIRAP and the superadiabatic STIRAP. We then show the experimental comparison between these two protocols for different shapes of the pulse envelopes in Sec.~\ref{sec:exp}. We  finally analyze the robustness of the protocols in Sec.~\ref{sec:robustness}, before we conclude in Sec.~\ref{sec:conclusion}.

\section{Traditional and superadiabatic STIRAP}
\label{sec:protocols}
For the sake of self-sufficiency, we brief summarize the theory describing both the traditional STIRAP and the two-photon-transition superadiabatic STIRAP (sa-STIRAP), i.e., the two population-transfer protocols whose performance we experimentally compare in the subsequent section. For details on their derivations, we refer to Refs.~\cite{gaubatz1990population,vitanov2017stimulated,giannelli2014three,petiziol2020superadiabatic} and references therein.

\subsection{Traditional STIRAP}
\label{sec:stirap}
The basic STIRAP process involves three states, connected by two time-dependent driving fields. Although these three states are oftentimes labeled $\{|1\rangle, |2\rangle, |3\rangle\}$ in the literature, we will label them $\{|0\rangle, |-1\rangle, |+1\rangle\}$ for later convenience.

\begin{figure}[t]
\flushleft\normalsize{(a)}\hspace{4cm}\normalsize{(b)}\\
\includegraphics[width=0.5\linewidth]{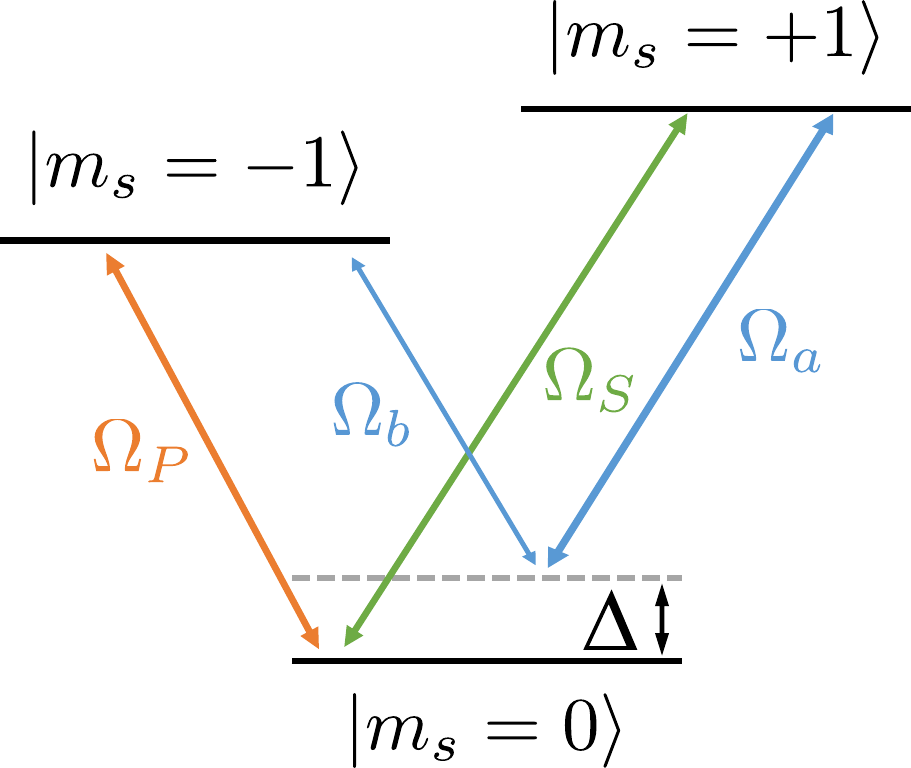}\hspace{0.5cm}
\includegraphics[width=0.43\linewidth]{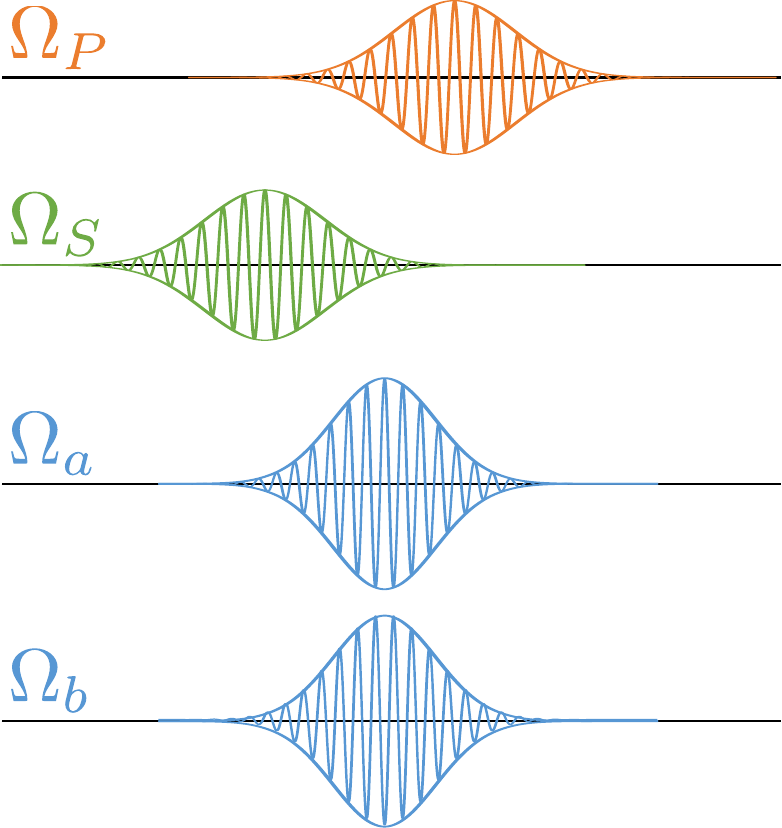}
\caption{(a) Level structure of the electronic ground-state triplet of the NV center spin in diamond. Using the intermediate state $|m_s=0\rangle$ population transfer from $|m_s = -1\rangle$ to $|m_s = +1\rangle$ is achieved with an adiabatic or superadiabatic protocol by applying two or four microwave driving fields, respectively. (b) Schematic time sequence of the four pulses: pump and Stokes pulses of the traditional STIRAP (top two) and superadiabatic corrections (bottom two).}
\label{fig:pulse_seq}
\end{figure}

The aim of STIRAP is a population transfer between the two states $|-1\rangle$ and $|+1\rangle$, which are indirectly coupled by using the intermediate state $|0\rangle$ without populating the latter. Typically, a pump field connects the initial state $|-1\rangle$ with $|0\rangle$, which is coupled to the final target quantum state $|+1\rangle$ by a Stokes field. Based on the energy relation of these three states, different level configurations might appear, such as ladder, $\Lambda$, or V-type systems. Without loss of generality, here we only consider a V-type configuration, since this is the level structure of the electronic ground-state triplet of the NV-center spin in diamond, which we will use later for experimental implementation. This level structure is depicted in Fig.~\ref{fig:pulse_seq}(a),  with the three states corresponding to the spin sublevels $\{|m_s = 0\rangle,|m_s = -1\rangle, |m_s = +1\rangle\}$, which is the reason why we have labeled the states in this fashion. 
The two transitions that can be driven are coupled by the pump field $\Omega_P$ and Stokes field $\Omega_S$. Under the two-photon resonance condition, the system Hamiltonian can be written as~\cite{gaubatz1990population,bergmann1998coherent}
\begin{equation}
\label{Eq:STIRAP_H}
    H_\mathrm{st}(t)=\frac{1}{2}\begin{pmatrix}
    0 & \Omega_P(t) & \Omega_S(t)\\
    \Omega_P(t) & 0 & 0\\
    \Omega_S(t) & 0 & 0\\
    \end{pmatrix}.
\end{equation}
Here, $\Omega_P(t)$ and $\Omega_S(t)$ in the STIRAP protocol represent the Rabi frequencies of the pump and Stokes, respectively, implemented via external microwave control fields. Counter-intuitively, the Stokes field is applied before the pump field, as shown by the time order in Fig.~\ref{fig:pulse_seq}(b), although there is initially no population in the two states it couples. The key mechanism behind the STIRAP protocol is the existence of a dark state, which is the instantaneous eigenstate of Hamiltonian~\eqref{Eq:STIRAP_H} with instantaneous eigenvalue zero and has the form
\begin{equation}
    |D(t)\rangle=\cos[\theta(t)]|-1\rangle-\sin[\theta(t)]|+1\rangle,
\end{equation}
with $\tan[\theta(t)]=\Omega_P(t)/\Omega_S(t)$. Since the pulses are counter-intuitively ordered, i.e.,  $\Omega_P(\ti)/\Omega_S(\ti) = 0$ and $\Omega_P(\tf)/\Omega_S(\tf) \approx \infty$, one finds $|D(t=\ti)\rangle=|-1\rangle$ and $|D(t=\tf)\rangle=|+1\rangle$, which means the dark state will coincide with the initial state $\ket{-1}$ at the initial time $\ti$ and with the target state $\ket{+1}$ at the final time $\tf$. Notice also that the dark state has a zero component of the intermediate state $\ket{0}$. Therefore, if the system is prepared in the initial state $\vert -1 \rangle$ and the evolution is adiabatic, the system will always remain in the dark state during the time evolution and transitions to other instantaneous eigenstates are suppressed~\cite{born1928beweis}. Consequently, at the final time, the system will be in the state $\vert D(\tf) \rangle = \ket{+1}$ without ever having populated the intermediate state $\ket{0}$. In summary, the STIRAP protocol allows us to efficiently transfer the population from the initial state $|-1\rangle$, which can easily be prepared in the NV center by optical ground-state polarization, to the final state $|+1\rangle$ by following the dark-state trajectory, as long as this is done adiabatically. The adiabatic condition is generally stated in its global form as $\Omega_0 \delta t\gg 10$~\cite{bergmann1998coherent}, where $\Omega_0$ is the peak Rabi frequency of both the pump and the Stokes pulse and $\delta t$ is the delay between them.

In general, different pulse shapes of the envelopes $\Omega_P(t)$ and $\Omega_S(t)$ might be designed and engineered to improve the adiabatic performance of the STIRAP protocol. The most commonly used paradigmatic scheme is the Gaussian pulse shape for the pump and Stokes fields, see Fig.~\ref{fig:stirap_pulse_shape}. However, due to the condition imposed by the adiabatic theorem to achieve high fidelity, the population transfer of the STIRAP protocol generally utilizes large average Rabi frequencies as well as possible long pulse durations, making it not suitable for many quantum tasks that require high efficiency in energy and time.

\subsection{Two-photon-transition superadiabatic STIRAP}
The critical disadvantages of the STIRAP protocol hinder its applicability in many scenarios where neither high pulse intensities or nor long transfer times are available. Therefore, recently, sa-STIRAP protocols have received much attention for overcoming this circumstance. Theoretically, it allows the controlled system to perfectly follow the instantaneous dark state by applying an additional Hamiltonian, with a possible transfer speed close to the quantum limit.

Following Ref.~\cite{giannelli2014three}, the two-photon-transition sa-STIRAP approach assumes a total control Hamiltonian of the three-level system that has the form
\begin{equation}
    H(t)=H_\mathrm{st}(t)+H_\mathrm{sa}(t).
\end{equation}
Specifically, the additional superadiabatic correction reads
\begin{equation}
    H_\mathrm{sa}(t)=i\sum_{n}\left[|\partial_t n(t)\rangle\langle n(t)|-\langle n(t)|\partial_t n(t)\rangle |n(t)\rangle\langle n(t)|\right],
\end{equation}
where the summation runs over the three states $|n(t)\rangle\in \{|D(t)\rangle,|B_-(t)\rangle,|B_+(t)\rangle\}$, i.e., the instantaneous eigenstates of the STIRAP Hamiltonian, $H_\mathrm{st}(t)$, in Eq.~\eqref{Eq:STIRAP_H}.

Experimentally, we add two additional microwave fields, denoted by $\Omega_a(t)$ and $\Omega_b(t)$, to the traditional STIRAP protocol, see Fig.~\ref{fig:pulse_seq}(b). Both of these two fields are detuned from the resonance of their respective $|0\rangle\leftrightarrow|\pm1\rangle$ transitions by the same amount, $\Delta$. The corresponding time-dependent Hamiltonian in the rotating frame has the form
\begin{equation}
    H_\mathrm{sa}=\frac{1}{2}\begin{pmatrix}
  0 & \Omega_a^\ast(t)e^{-i\Delta t} & \Omega_b(t)e^{-i\Delta t} \\
  \Omega_a(t)e^{i\Delta t}  & 0 & 0\\
  \Omega_b^\ast(t)e^{i\Delta t}  & 0 & 0
    \end{pmatrix}.
\end{equation}
Here, the two Rabi frequencies $\Omega_a(t)$ and $\Omega_b(t)$ are chosen as 
\begin{gather}
    \label{eq:omega_a}
    \Omega_a(t)=e^{i \frac{\pi}{2}}\sqrt{2\Delta|\Omega_d(t)|},\\
    \label{eq:omega_b}
    \Omega_b(t)=\sqrt{2\Delta|\Omega_d(t)|}.
\end{gather}
where the effective two-photon pulse has the form 
\begin{equation}
\label{eq:Omega_d}
  \Omega_d(t)= -2i \frac{\dot{\Omega}_P(t)\Omega_S(t) - \Omega_P(t)\dot{\Omega}_S(t)}{\Omega_P(t)^2 + \Omega_S(t)^2}.
\end{equation}
Details on how this connection to the shape of the STIRAP pulses $\Omega_S(t)$ and $\Omega_p(t)$ arises, can be found, for example, in Refs.~\cite{giannelli2014three,petiziol2020superadiabatic}. The shape of the two pulses $\Omega_{a}$ and $\Omega_{b}$ then directly follow from the equations above. Note that the conditions $|\Delta|\gg|\Omega_{a}|,|\Omega_{b}|$ has to be fulfilled.

\section{Experimental realization for different pulse shapes}
\label{sec:exp}
Details on our experiment setup have been given, among others, in Ref.~\cite{gong2023accelerated}. As mentioned before, in the experiment, we use the ground-state triplet of a single NV-center spin in diamond to compare the performance of the sa-STIRAP to the traditional one. We do so for different envelope shapes of the pump and Stokes pulses.

\subsection{Gaussian pulses}
\label{sec:gaussian}
\begin{figure}[tb]
    \centering
    \includegraphics[width=\linewidth]{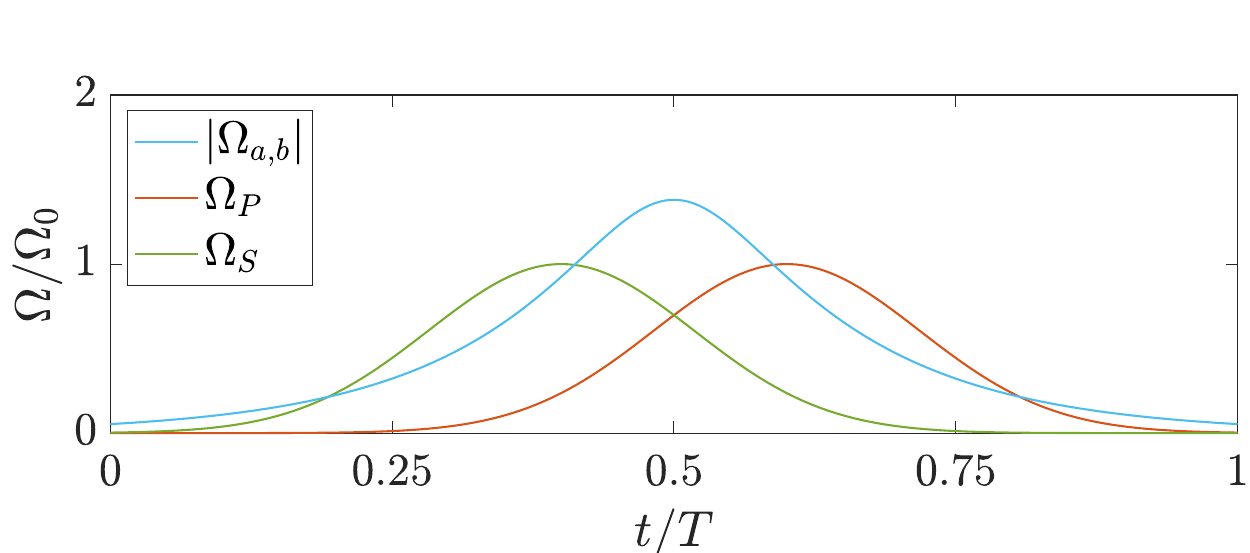}
    \caption{The two Raman pulses in the Gaussian STIRAP scheme (orange and green) and the two superadiabatic correction pulses (blue). The green and orange lines correspond to the partially-overlapping control pulses $\Omega_S(t)$ and $\Omega_P(t)$, respectively. Since the two additional fields $\Omega_a(t)$ and $\Omega_b(t)$ have the same magnitude and only different phases, they are represented by a single blue line.}
    \label{fig:stirap_pulse_shape}
\end{figure}

We begin by checking the most widely used envelope shape, i.e., the STIRAP protocol with Gaussian $\Omega_S(t)$ and $\Omega_P(t)$, see Fig.~\ref{fig:stirap_pulse_shape}. 
By denoting the total evolution time by $T$ and setting the pulses to be centered at $T/2\pm\delta t$, i.e., the second pulse is delayed by $2\delta t$ with respect to the first one, the two pulses can be expressed in the form
\begin{gather}
\label{eq:gaussian1}
\Omega_S (t)=\Omega_0\exp\left[-\frac{1}{\sigma^2}\left(t-\frac{T}{2}+\delta t\right)^2\right],\\
\label{eq:gaussian2}
\Omega_P (t)=\Omega_0\exp\left[-\frac{1}{\sigma^2}\left(t-\frac{T}{2}-\delta t\right)^2\right], 
\end{gather}
where $\sigma$ represents the half-width at half-maximum of the envelope. 
 
With these Raman-pulse envelopes the magnitude of the effective two-photon pulse can be derived from Eq.~\eqref{eq:Omega_d}, which yields
 \begin{equation}
    |\Omega_d(t)|=\frac{4\delta t}{\sigma^2}\left\{\cosh\left[\frac{4\delta t}{\sigma^2}\left(t-\frac{T}{2}\right)\right]\right\}^{-1}.
\end{equation}
 According to Eqs.~\eqref{eq:omega_a} and~\eqref{eq:omega_b}, the two superadiabatic corrections~\eqref{eq:omega_a} and~\eqref{eq:omega_b} that have to be applied on top of the STIRAP in order to enhance the transfer efficiency thereby have the form
\begin{gather}
    \Omega_a(t)=e^{i\frac{\pi}{2}}\sqrt{\frac{8\Delta\delta t}{\sigma^2}}\left\{\cosh\left[\frac{4\delta t}{\sigma^2}\left(t-\frac{T}{2}\right)\right]\right\}^{-1/2},\\
    \Omega_b(t)=\sqrt{\frac{8\Delta\delta t}{\sigma^2}}\left\{\cosh\left[\frac{4\delta t}{\sigma^2}\left(t-\frac{T}{2}\right)\right]\right\}^{-1/2}.
\end{gather}

In our experiments, we set $\Omega_0/2\pi=2$ MHz, $\delta t=T/10$, and $\sigma=T/6$. The strict conditions on the detuning $\Delta$ for the two-photon transition would require $|\Delta|\gg 8\delta t/\sigma^2=28.8/T$. However, when combining STIRAP with the two-photon transition, we have to apply four microwave fields with the amplitudes $\Omega_S$, $\Omega_P$, $\Omega_a$, and $\Omega_b$ but the total output power of the arbitrary waveform generator is limited, such that we may not choose them arbitrarily large. In the experiment, the Rabi frequency of the combined microwave fields should be below $10$~MHz. Since we chose $\Omega_0/2\pi=2$~MHz, at most 3 MHz remain for $\Omega_a$ and $\Omega_b$ each. In principle, $\Delta$ should be much larger than both these amplitudes, but due to the narrow parameter range we are facing we choose $\Delta/2\pi=3$~MHz. This leads to $|\Omega_a|/2\pi=|\Omega_b|/2\pi=2.6221$~MHz. Although the strict condition $|\Delta|\gg|\Omega_{a}|,|\Omega_{b}|$ is thereby not fulfilled, we nevertheless find that the protocol works.

\begin{figure*}
    \centering
    \flushleft\normalsize{(a)}\hspace{5.6cm}\normalsize{(b)}\hspace{5.6cm}\normalsize{(c)}\\
    \includegraphics[width=0.33\linewidth]{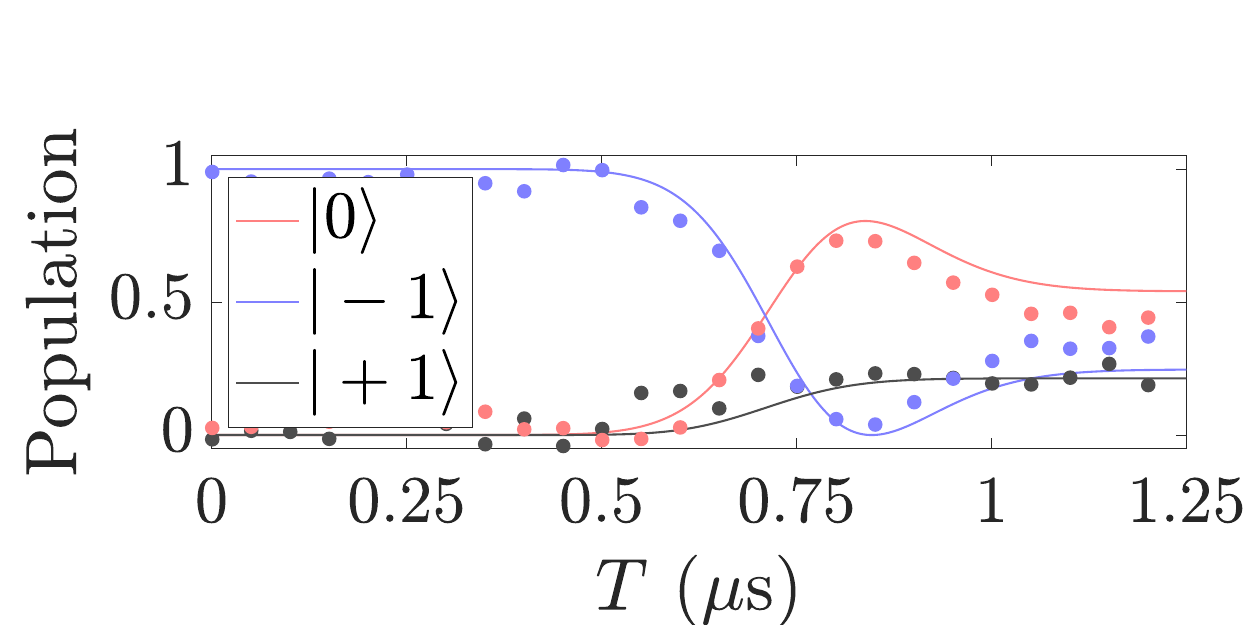}
    \includegraphics[width=0.33\linewidth]{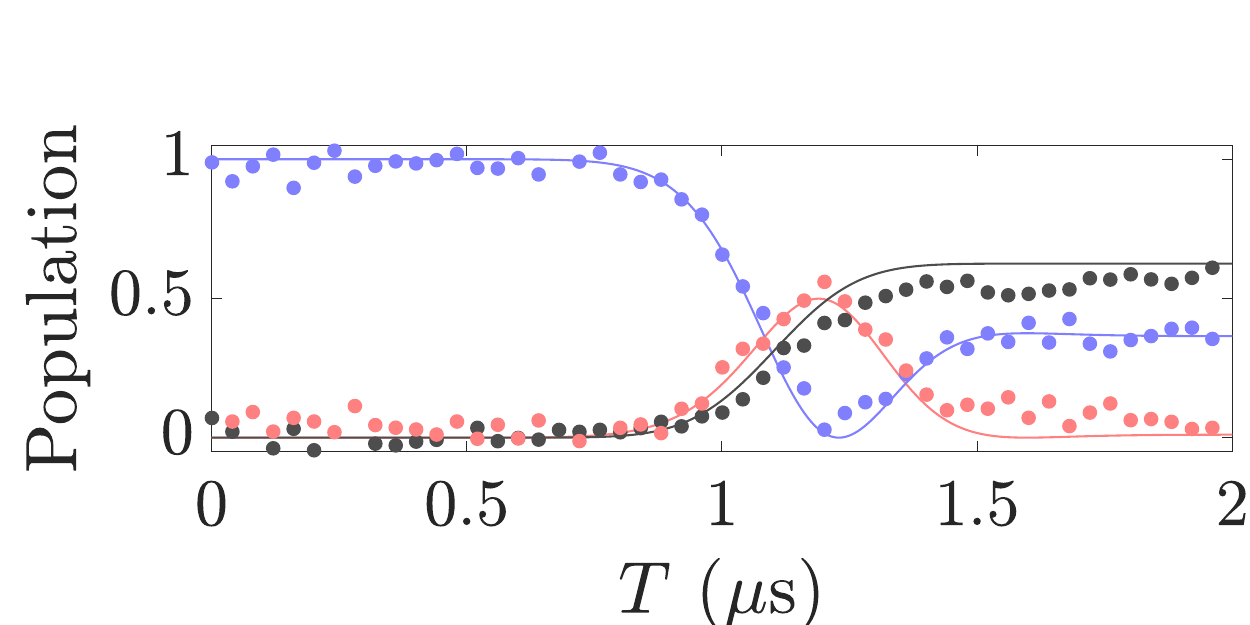}
    \includegraphics[width=0.33\linewidth]{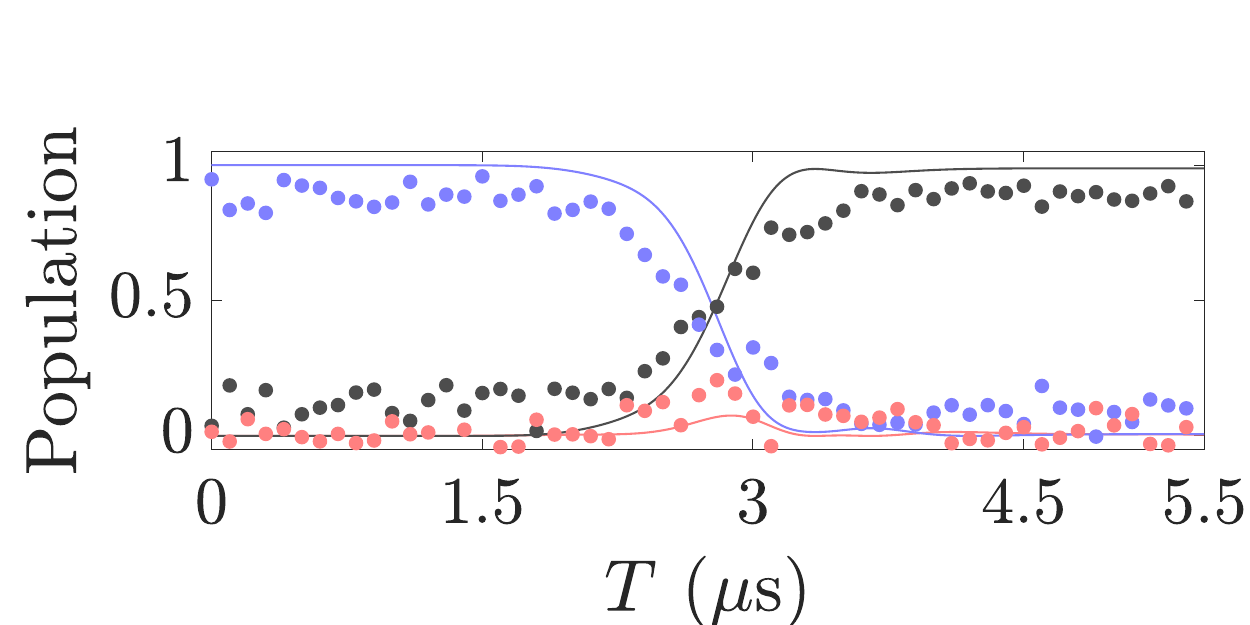}\vspace{-0.5cm}
    
    \flushleft\normalsize{(d)}\hspace{5.6cm}\normalsize{(e)}\hspace{5.6cm}\normalsize{(f)}\\
    \includegraphics[width=0.33\linewidth]{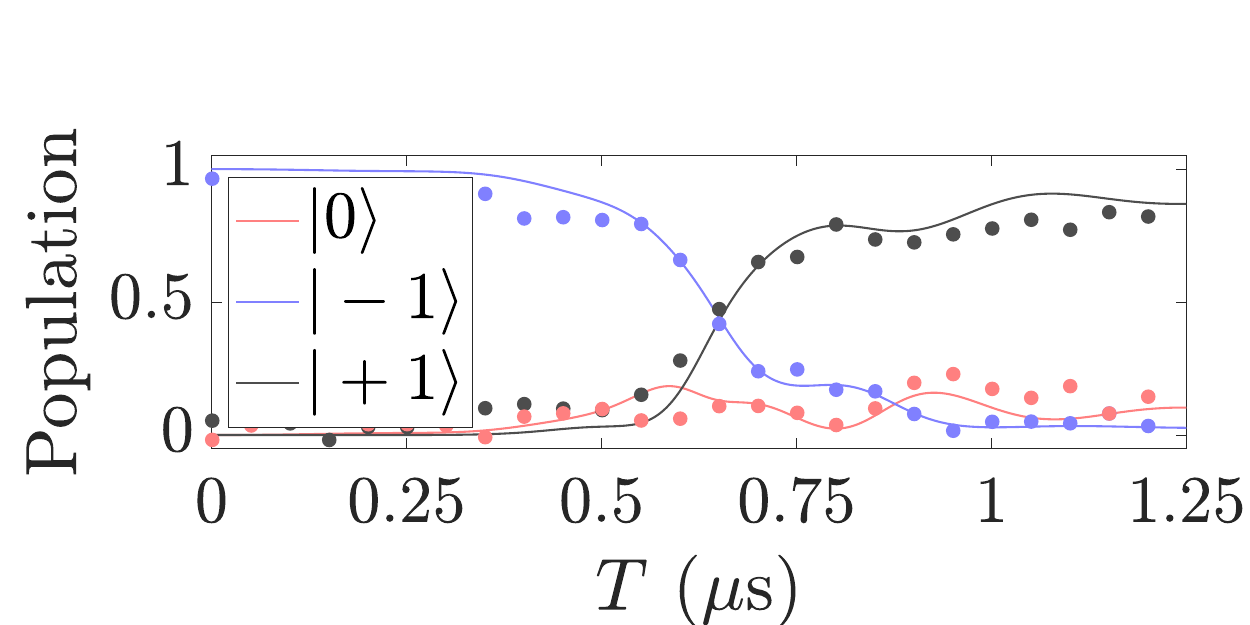}
    \includegraphics[width=0.33\linewidth]{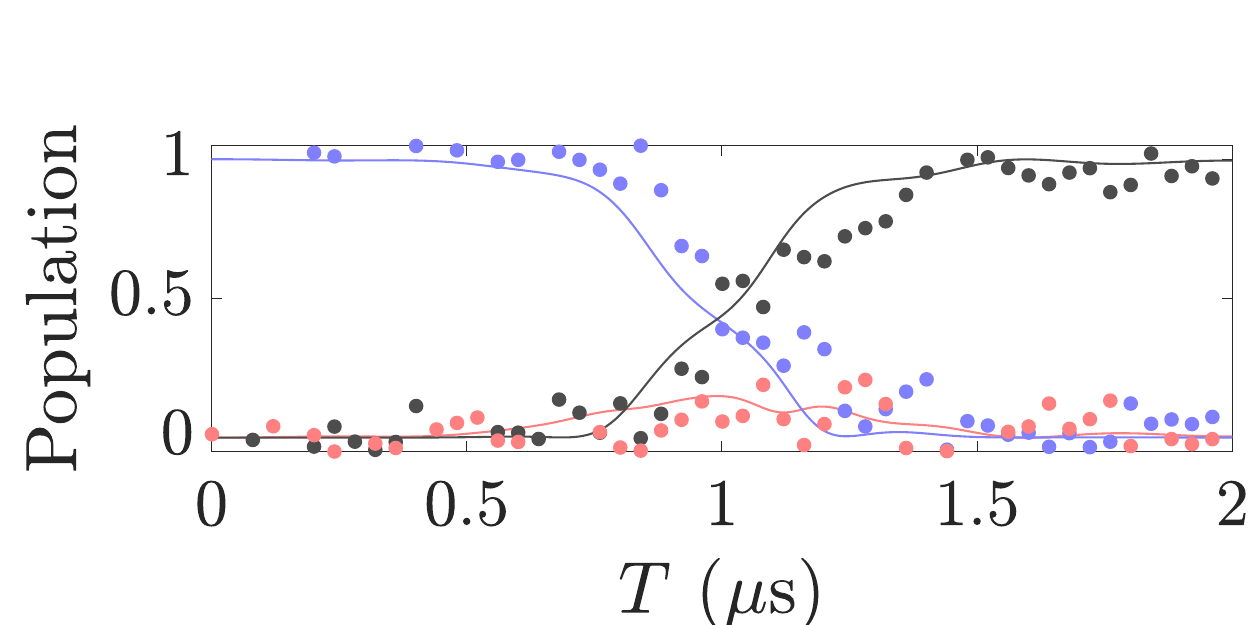}
    \includegraphics[width=0.33\linewidth]{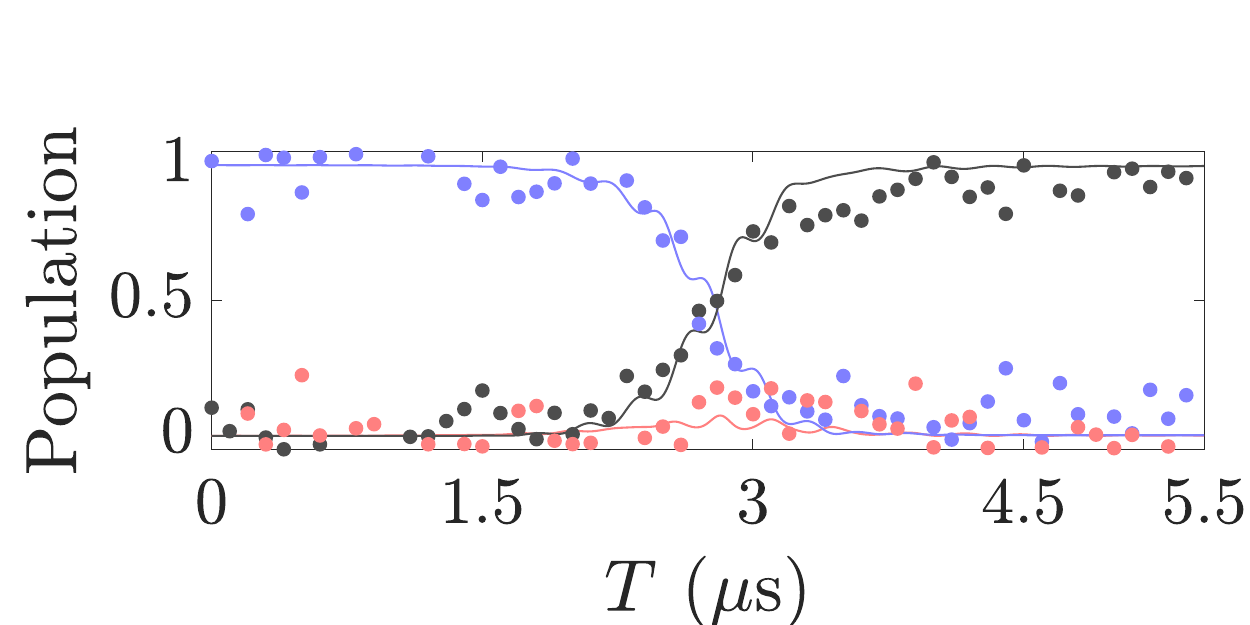}
	\caption{Evolution of the population of the three electronic states of the NV-center ground-state triplet during the STIRAP (upper row) and sa-STIRAP (lower row) protocol with Gaussian pulse envelopes. The parameters are $\Omega_0/2\pi=2$ MHz, $\Delta/2\pi=3$ MHz, $\delta t=T/10$, and $\sigma=T/6$. The three columns (from left to right) correspond to the three values $T=1.25$~$\mu$s, $T=2$~$\mu$s, and $T=5.5$~$\mu$s. Markers show experimental results whereas solid lines are numerical simulations.}
	\label{fig:evolution_gaussian}
\end{figure*}

For the set of parameters described above, Fig.~\ref{fig:evolution_gaussian} shows a comparison of STIRAP and sa-STIRAP for three different values of the total pulse duration $T$.
Here, the upper row shows the STIRAP whereas the lower row shows the sa-STIRAP. The three columns correspond to $T=1.25$~$\mu$s, $T=2$~$\mu$s, and $T=5.5$~$\mu$s. The figure shows the population of all three states, the intermediate state $|0\rangle$ in red, the initial state $|-1\rangle$ in blue, and the target state $|+1\rangle$ in black. Markers show experimental results and solid lines are numerical simulations. One clearly sees that the sa-STIRAP out performs the traditional STIRAP in all three cases. Even for the relatively short duration $1.25$~$\mu$s the sa-STIRAP is able to achieve a high transfer efficiency, as given by the population of the target state, while the traditional STIRAP shows a weak performance and is only able to keep up with the superadiabatic protocol at much longer durations, such as $5.5$~$\mu$s in the last column. It is also important to note that the sa-STIRAP shows an appreciably lower population of the intermediate state throughout all experiments, while the STIRAP protocol populates the intermediate state for short transfer times $T$, because in this regime the adiabatic condition is not fulfilled.

\begin{figure}[tb]
\centering
\includegraphics[width=\linewidth]{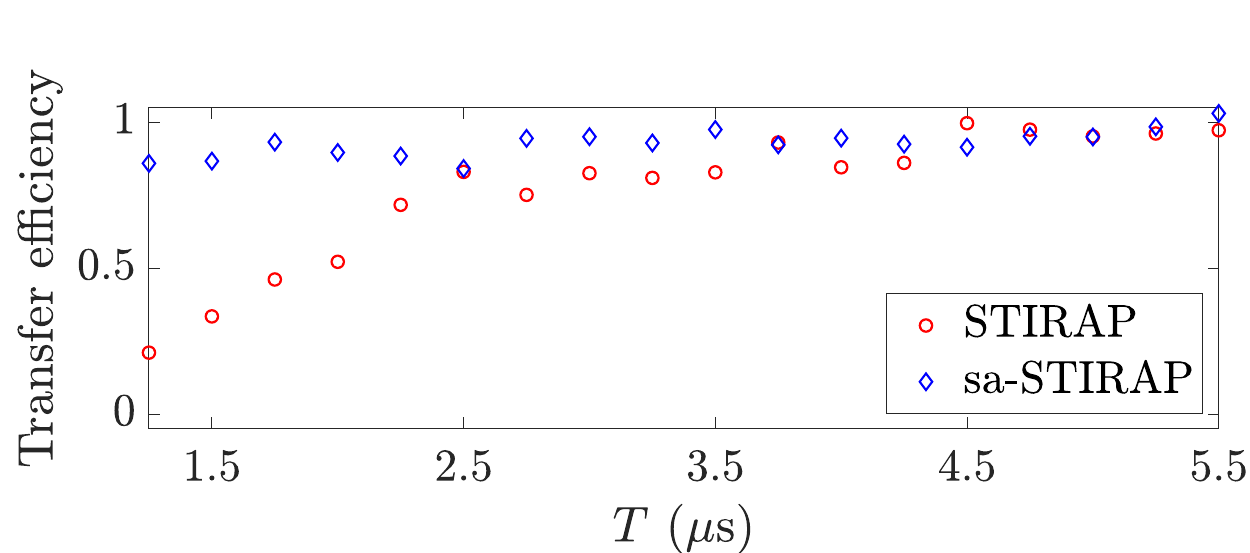}
\caption{Dependence of the transfer efficiency in a three-level system on the protocol duration $T$. STIRAP with Gaussian pulses (red circles) and the corresponding sa-STIRAP (blue diamonds). The parameters are the same as in Fig.~\ref{fig:evolution_gaussian}.}
\label{fig:eff_comparison}
\end{figure}

Let us have a closer look at the fact that the two additional fields resulted in an increased transfer efficiency for the three parameters we chose above. In Fig.~\ref{fig:eff_comparison}, we emphasize this fact by showing the state-transfer efficiency for both traditional STIRAP and sa-STIRAP as a function of the evolution time $T$. 
One sees that the latter protocol shows an enhanced transfer efficiency throughout the whole range of total evolution times we chose to measure, again showing the speed up that can be attained by adding the superadiabatic corrections.

\subsection{Exponential pulses}
\begin{figure}[tb]
\flushleft\normalsize{(a)}\hspace{-2.4ex}\includegraphics[width=\linewidth]{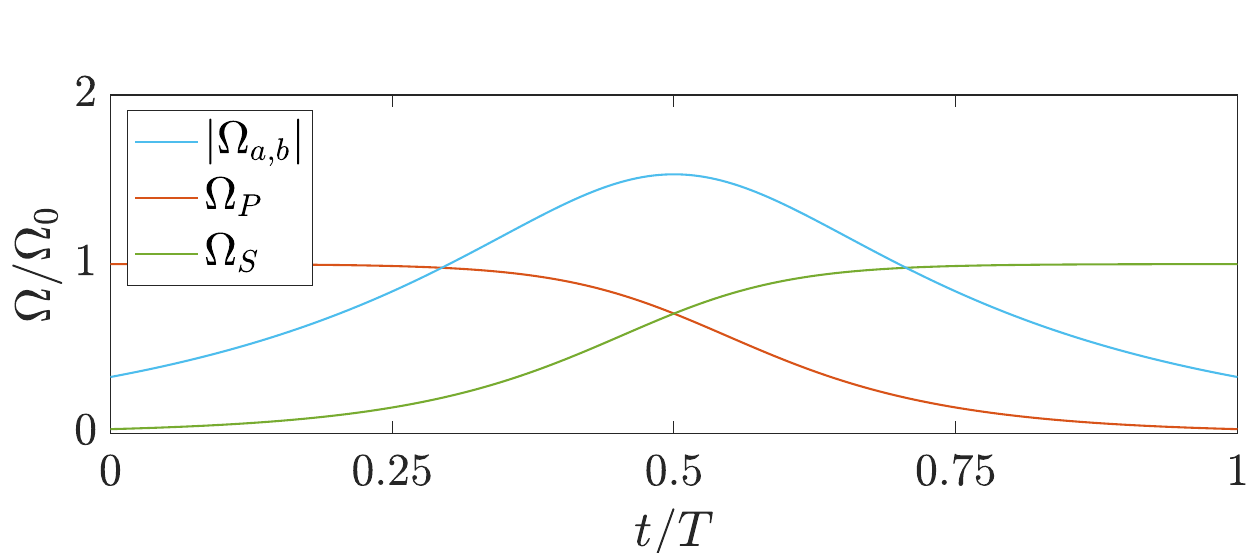}\vspace{-3ex}
\flushleft\normalsize{(b)}\hspace{-2.4ex}\includegraphics[width=\linewidth]{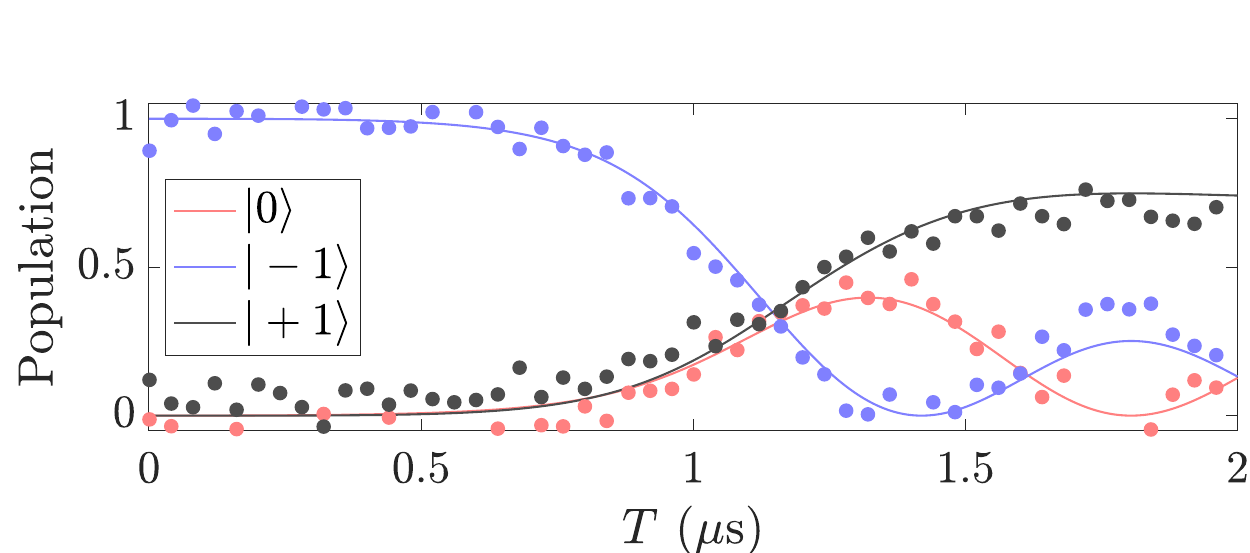}\vspace{-3ex}
\flushleft\normalsize{(c)}\hspace{-2.4ex}\includegraphics[width=\linewidth]{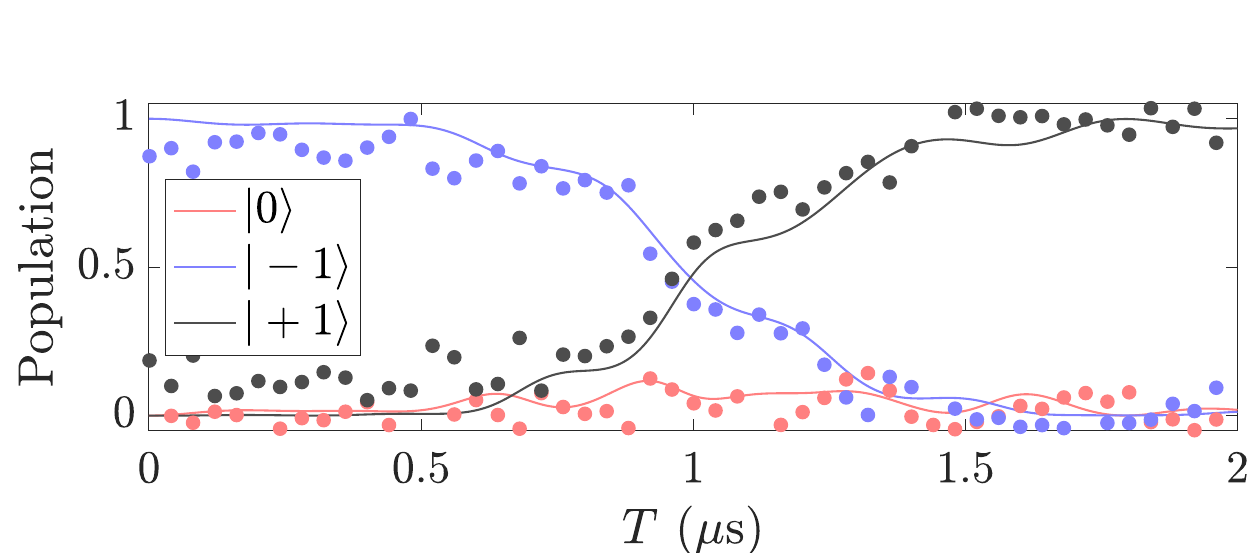}
\caption{(a) The two Raman pulses of the STIRAP scheme (orange and green) with exponential envelopes and the two corresponding superadiabatic correction pulses (blue). For these pulse shapes the evolution of the population of the three states is shown in (b) for the STIRAP and in (c) for the  sa-STIRAP protocols. Markers show experimental results and solid lines are numerical simulations. The parameters are $\Omega_0/2\pi = 1.2$~MHz, $\Delta/2\pi = 3$~MHz, $\sigma = T/15$, and $T = 2$~$\mu$s.}
\label{fig:exponential_pulse}
\end{figure}

Let us also show that an enhanced performance of the sa-STIRAP protocol is attained also with other pulse shapes. Therefore, as a second example for possible choices of the STIRAP pulses we choose the exponential shapes~\cite{laine1996exponential,giannelli2014three} given by
\begin{gather}
\label{eq:exp1}
\Omega_S (t)=\Omega_0\left\{1+\exp\left[-\frac{1}{\sigma}\left(t-\frac{T}{2}\right)\right]\right\}^{-1/2},\\
\label{eq:exp2}
\Omega_P (t)=\Omega_0\left\{1+\exp\left[\frac{1}{\sigma}\left(t-\frac{T}{2}\right)\right]\right\}^{-1/2}.
\end{gather}
These envelope shapes of the control pulses are depicted in Fig.~\ref{fig:exponential_pulse}(a).
One can, again, use Eq.~\eqref{eq:Omega_d} to derive the shape of effective two-photon-transition pulse, which yields
 \begin{equation}
    |\Omega_d(t)|=\frac{1}{2\sigma}\left\{\cosh\left[\frac{1}{2\sigma}\left(t-\frac{T}{2}\right)\right]\right\}^{-1}.
\end{equation}
In a next step, one can then use Eqs.~\eqref{eq:omega_a} and~\eqref{eq:omega_b} to obtain the shape of the two superadiabatic-correction pulse envelopes
\begin{gather}
    \Omega_a(t)=e^{i\frac{\pi}{2}}\sqrt{\frac{\Delta}{\sigma}}\left\{\cosh\left[\frac{1}{2\sigma}\left(t-\frac{T}{2}\right)\right]\right\}^{-1/2},\\
    \Omega_b(t)=\sqrt{\frac{\Delta}{\sigma}}\left\{\cosh\left[\frac{1}{2\sigma}\left(t-\frac{T}{2}\right)\right]\right\}^{-1/2}.
\end{gather}

In order to demonstrate that the sa-STIRAP also performs better in this case, we chose the total duration $T=2$~$\mu$s, where the advantage over traditional STIRAP using Gaussian pulses was shown in the central column of Fig.~\ref{fig:evolution_gaussian}. We furthermore chose $\Omega_0/2\pi = 1.2$~MHz and $\Delta/2\pi = 3$~MHz, as well as the width $\sigma = T/15$. The results of STIRAP and sa-STIRAP are shown in Figs.~\ref{fig:exponential_pulse}(b) and~\ref{fig:exponential_pulse}(c), where the population of all three states is shown as markers for experimental results and as solid lines for numerical simulations. Even if the superadiabatic corrections are not fully applied during the protocol duration [seen in Fig.~\ref{fig:exponential_pulse}(a), for example, where $\Omega_a(t)$ and $\Omega_b(t)$ do not reach zero at the beginning and at the end of the protocol], also for this shape of the pulse envelopes the sa-STIRAP leads to a clearly visible improvement of the state-transfer efficiency as well as a decreased population of the intermediate state. This also shows that the superadiabatic protocol is robust against fluctuations of the corrections $\Omega_a(t)$ and $\Omega_b(t)$.

\subsection{Trigonometric pulses}
\begin{figure}[tb]
\flushleft\normalsize{(a)}\hspace{-2.4ex}\includegraphics[width=\linewidth]{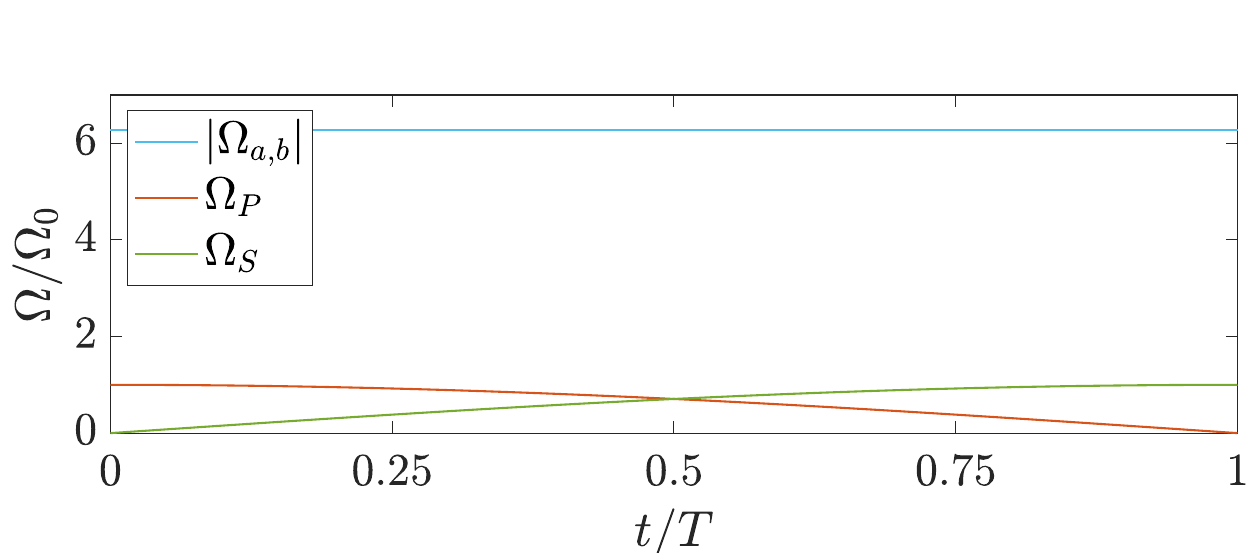}\vspace{-3ex}
\flushleft\normalsize{(b)}\hspace{-2.4ex}\includegraphics[width=\linewidth]{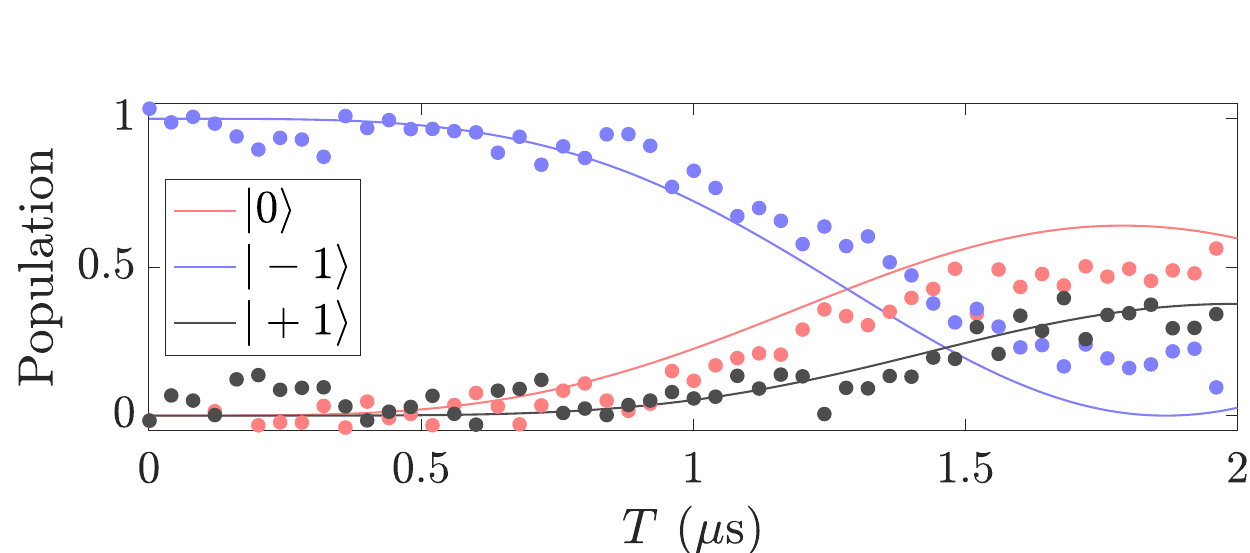}\vspace{-3ex}
\flushleft\normalsize{(c)}\hspace{-2.4ex}\includegraphics[width=\linewidth]{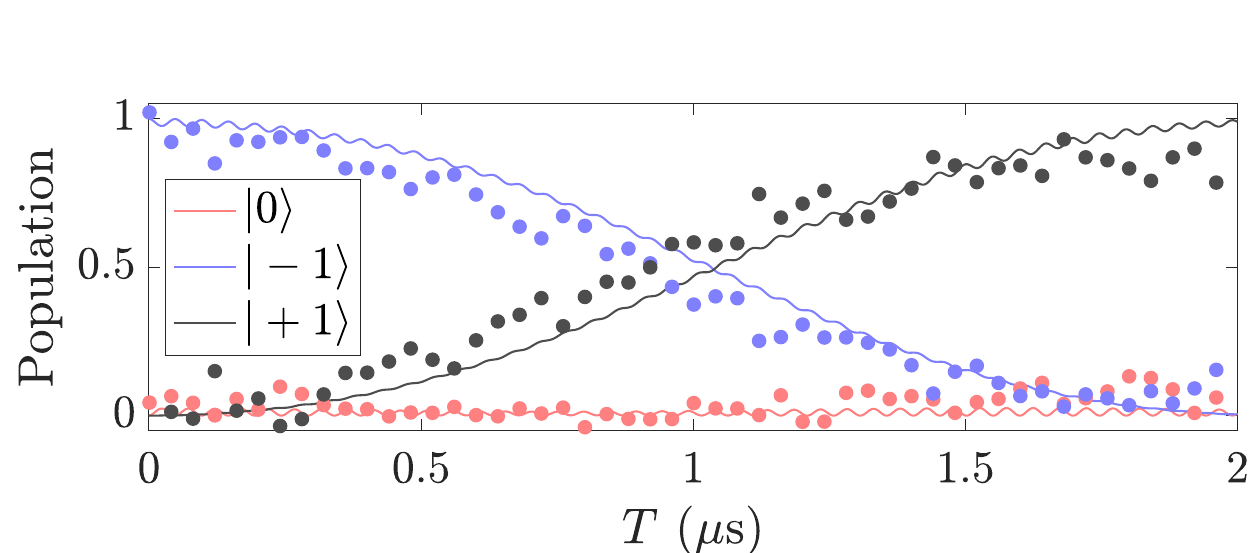}
\caption{(a) The two Raman pulses of the STIRAP scheme (orange and green) with trigonometric envelopes and the two corresponding superadiabatic correction pulses (blue). For these pulse shapes the evolution of the population of the three states is shown in (b) for the STIRAP and in (c) for the  sa-STIRAP protocols. Markers show experimental results and solid lines are numerical simulations. The parameters are $\Omega_0/2\pi = 0.5$~MHz, $\Delta/2\pi = 20$~MHz, and $T = 2$~$\mu$s.}
\label{fig:trigonometric_pulse}
\end{figure}

As a third and final example for different shapes of the control-pulse envelopes we choose the trigonometric pulse shapes~\cite{laine1996exponential,chen2012trigonometric,giannelli2014three}
\begin{gather}
    \Omega_S(t) = \Omega_0 \sin\left(\frac{\pi t}{2 T}\right),\\
    \Omega_P(t) = \Omega_0 \cos\left(\frac{\pi t}{2 T}\right),
\end{gather}
which are shown in Fig.~\ref{fig:trigonometric_pulse}(a). 
Upon inserting this into Eq.~\eqref{eq:Omega_d} one retrieves a magnitude of the effective two-photon-transition pulse given by the constant
\begin{equation}
    |\Omega_d(t)|= \frac{\pi}{T}.
\end{equation}
This, in turn, leads to the constant superadiabatic corrections 
\begin{gather}
    \Omega_a(t)= e^{i \frac{\pi}{2}}\sqrt{\frac{2 \pi \Delta}{T}},\\
    \Omega_b(t)=\sqrt{\frac{2 \pi \Delta}{T}}.
\end{gather}

For the experimental comparison, we chose the same pulse duration as in the previous example, i.e, $T = 2$~$\mu$s. Furthermore, we have fixed the Rabi frequency to  be $\Omega_0/2\pi=0.5$~MHz and the detuning to be $\Delta/2\pi=20$~MHz. The results for STIRAP and sa-STIRAP are shown in Figs.~\ref{fig:trigonometric_pulse}(b) and~~\ref{fig:trigonometric_pulse}(c), respectively. As before, we show the population of all three states as markers for the experimental results and as solid lines for numerical simulations. Also in this example for possible pulse shapes the sa-STIRAP shows a tremendous improvement in the transfer efficiency as well as vanishingly small population of the intermediate state during the transfer.

\section{Robustness against pulse imperfections}
\label{sec:robustness}
In the previous section it became clear that the sa-STIRAP can achieve very high state-transfer efficiency for pulse durations where traditional STIRAP is unable to do so while keeping the population of the intermediate state very low throughout the whole transfer. Apart from this advantage of the superadiabatic protocol, let us investigate the impact of the superadiabatic corrections on the robustness against pulses imperfections.

\begin{figure}[tb]
    \flushleft\normalsize{(a)}\hspace{4cm}\normalsize{(b)}\\ 
    \includegraphics[width=0.495\linewidth]{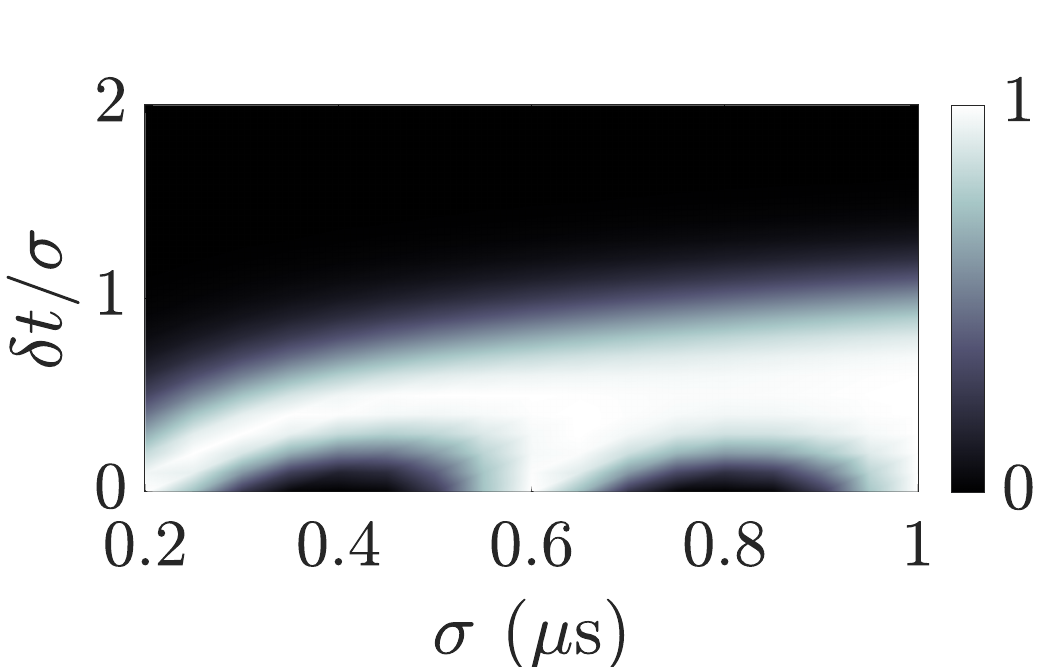}
    \includegraphics[width=0.495\linewidth]{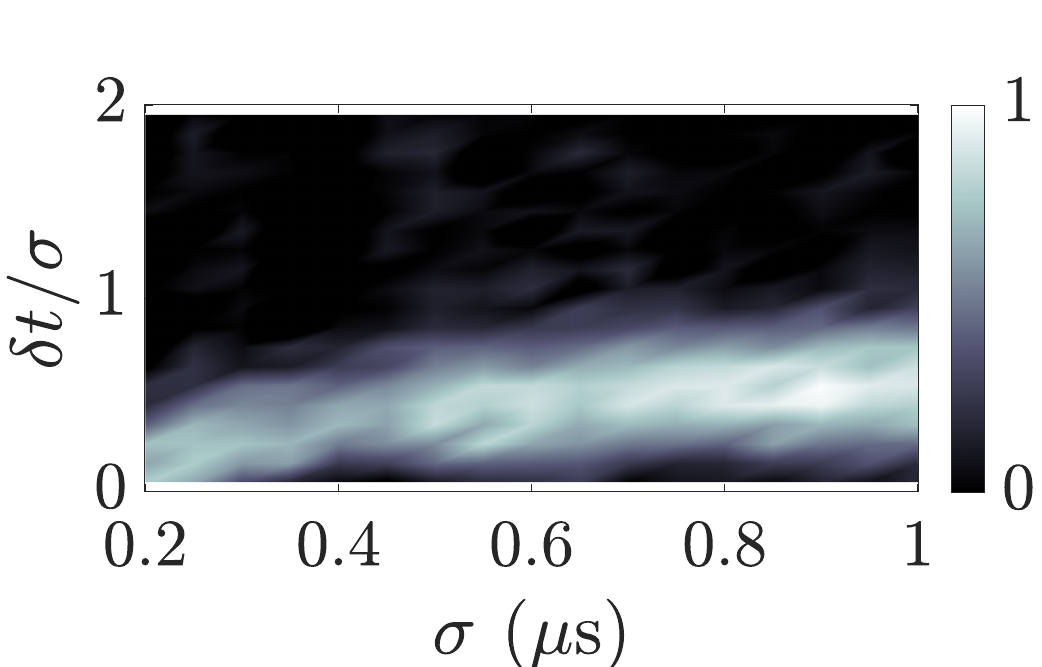}\vspace{-0.5cm}
    
    \flushleft\normalsize{(c)}\hspace{4cm}\normalsize{(d)}\\ 
    \includegraphics[width=0.495\linewidth]{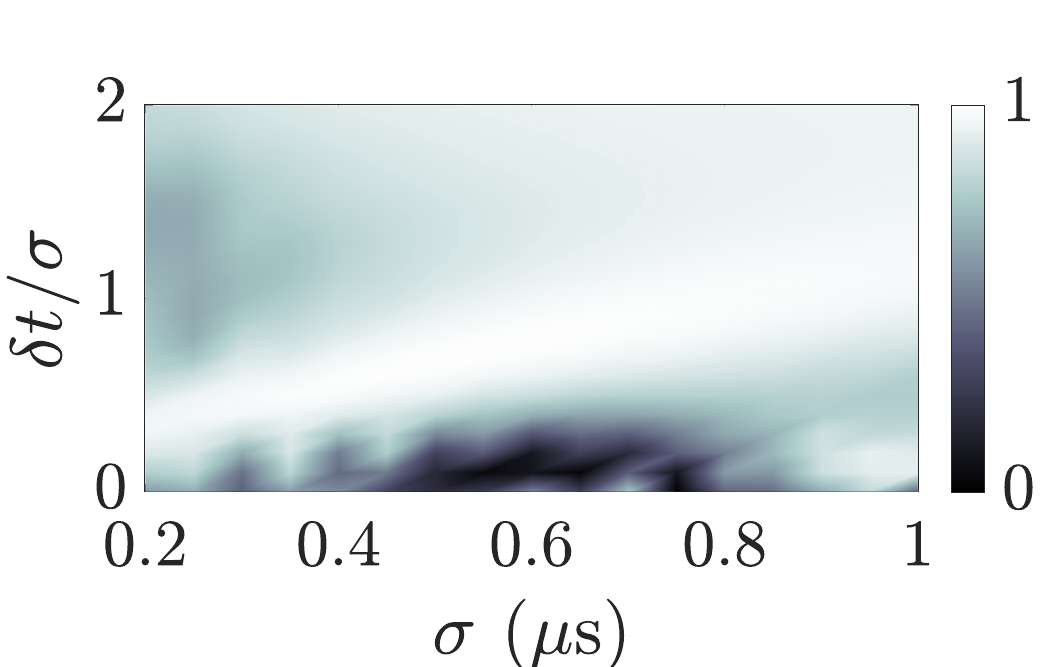}
    \includegraphics[width=0.495\linewidth]{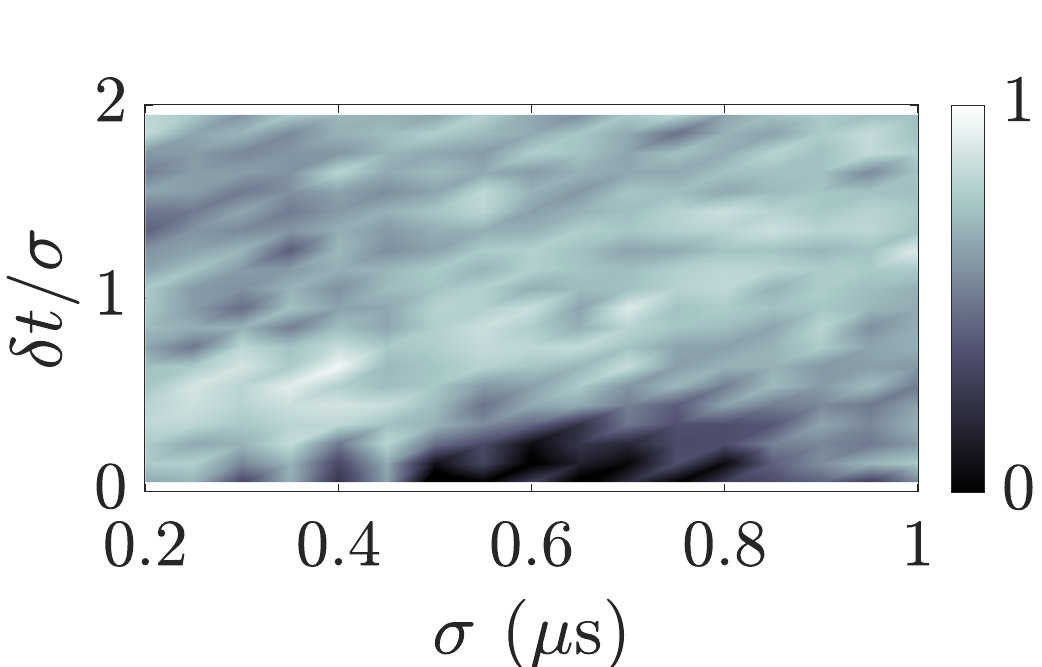}
    \caption{Comparison of the robustness against pulse imperfections between STIRAP (upper row) and sa-STIRAP (lower row) for Gaussian pulse shapes. Numerical simulations are shown in the left column and experimental results in the right column. We show the transfer efficiency in dependence of $\sigma$ (varying from 0.2~$\mu$s to 1~$\mu$s) and $\delta t$ (varying from 0 to $2\sigma$). For all points, the pulses for the superadiabatic correction are the ones optimized for the central point, viz., $\sigma=0.6$~$\mu$s and $\delta t = 0.6$~$\mu$s. The Rabi frequency and detuning are given by $\Omega_0/2\pi=2$~MHz and $\Delta/2\pi=3$~MHz, respectively.}
\label{fig:robustness}
\end{figure}

In fact, STIRAP is known to be relatively robustness against such pulses imperfections. This can be seen in the upper row of Fig.~\ref{fig:robustness}, which shows the transfer efficiency in dependence of the half-width at half-maximum of the Gaussian and the time delay between the Stokes and pump pulses. 
Here, the left column depicts numerical simulations and the right column shows experimental results. Explicitly, the parameter range is $0.2\ \mu\textrm{s}\leq \sigma\leq 1\ \mu\textrm{s}$ and $0\leq\delta t\leq 2\sigma$. For every pair of parameters, the duration of the protocol is given by $T=6\sigma+2\delta t$. In the upper row of Fig.~\ref{fig:robustness}, one sees relatively big areas that are bright in both numerical simulations [Fig.~\ref{fig:robustness}(a)] and experimental results [Fig.~\ref{fig:robustness}(b)], which correspond to Gaussian STIRAP pulses leading to a high-fidelity population transfer.

What we now show is that adding fixed superadiabatic correction pulses significantly increases this high-fidelity area. For all STIRAP pulses in the parameter range we add the same superadiabatic corrections that are optimized for the parameters $\sigma=\delta t=0.6$~$\mu$s, i.e., the central point of the parameter range. The corresponding results are shown in the lower row of Fig.~\ref{fig:robustness} for numerical simulations [Fig.~\ref{fig:robustness}(c)] and experimental results [Fig.~\ref{fig:robustness}(d)]. One finds that the original bright area of the STIRAP has been greatly extended by the sa-STIRAP correction, although the added correction is only the correct one for the central parameter pair. This shows another one of the tremendous advantages of the sa-STIRAP.

\section{Conclusions}
\label{sec:conclusion}
We have implemented a sa-STIRAP protocol based on a two-photon transition in a solid-state spin system. In order to benchmark its performance we have compared it with the traditional STIRAP protocol. The sa-STIRAP protocol employs two additional microwave driving pulses and is thereby able to improve population-transfer efficiency significantly. 

We have demonstrated this for three different shapes of the STIRAP pulse envelopes, including the widely-used Gaussian shape, an exponential shape, and a trigonometric shape. For all these pulses the transfer efficiencies are greatly improved by adding the superadiabatic corrections. This allows for a much faster population transfer without the necessity to overly increase the pulse intensities. We have also seen that throughout the transfer process, the population of the intermediate state is reduced by far, as compared to the traditional STIRAP. Furthermore, we have shown that adding the two superadiabatic corrections can increase the transfer fidelity even when the correction pulses are not optimized for the STIRAP pulses, making sa-STIRAP a very robust protocol. These advantages make the sa-STIRAP a great alternative to the traditional STIRAP.

\begin{acknowledgments}
M. G., M. Y., Y. C., W. C., N. W., and J. M. are supported by the National Natural Science Foundation of China (Grants No.~12161141011, No.~12074131), the National Key R$\&$D Program of China (Grant No. 2018YFA0306600), the Shanghai Key Laboratory of Magnetic Resonance (East China Normal University). Y. C. is also supported by a China Postdoctoral Science Foundation Grant (No. 2022M721256). R. B. is thankful to S. Schein for helpful discussions and comments. L. G. acknowledges the QuantERA grant SiUCs (Grant No. 731473) and the PNRR MUR project PE0000023-NQSTI.
\end{acknowledgments}


%

\end{document}